\documentclass[letterpaper]{article} 
\usepackage{aaai24}  
\usepackage{times}  
\usepackage{helvet}  
\usepackage{courier}  
\usepackage[hyphens]{url}  
\usepackage{graphicx} 
\usepackage{natbib}  
\usepackage{caption} 
\usepackage{algorithm}
\usepackage{algorithmic}

\usepackage{newfloat}
\usepackage{listings}
\DeclareCaptionStyle{ruled}{labelfont=normalfont,labelsep=colon,strut=off} 
\floatstyle{ruled}
\newfloat{listing}{tb}{lst}{}
\floatname{listing}{Listing}
\usepackage{soul}

\usepackage{array,booktabs}
\newcolumntype{C}{>{$\displaystyle}c<{$}}

\usepackage[dvipsnames]{xcolor}

\title{High Epsilon Synthetic Data Vulnerabilities in MST and PrivBayes}

\author{
    Steven Golob\textsuperscript{\rm 1},
    Sikha Pentyala\textsuperscript{\rm 1},
    Anuar Maratkhan\textsuperscript{\rm 1},
    Martine De Cock\textsuperscript{\rm 1}\thanks{Guest Professor at Ghent University}
}
\affiliations{
    \textsuperscript{\rm 1}University of Washington Tacoma\\
    \{golobs, sikha, anuar, mdecock\}@uw.edu
}

\begin{document}

\maketitle

\begin{abstract}

Synthetic data generation (SDG) has become increasingly popular as a privacy-enhancing technology. It aims to maintain important statistical properties of its underlying training data, while excluding any personally identifiable information. There have been a whole host of SDG algorithms developed in recent years to improve and balance both of these aims. Many of these algorithms provide robust differential privacy guarantees. 

However, we show here that if the differential privacy parameter $\varepsilon$ is set too high, then unambiguous privacy leakage can result. We show this by conducting a novel membership inference attack (MIA) on two state-of-the-art differentially private SDG algorithms: MST and PrivBayes. Our work suggests that there are vulnerabilities in these generators not previously seen, and that future work to strengthen their privacy is advisable. 

We present the heuristic for our MIA here. It assumes knowledge of auxiliary ``population'' data, and also assumes knowledge of which SDG algorithm was used. We use this information to adapt the recent DOMIAS MIA uniquely to MST and PrivBayes. Our approach went on to win the SNAKE challenge in November 2023.

\end{abstract}

\section{Introduction}

The development of techniques for synthetically generating data has burgeoned in recent years. First in Europe with the GDPR\footnote{European General Data Protection Regulation\\ \url{https://gdpr-info.eu/}} in 2018, and now in the United States\footnote{The AI Bill of Rights, unveiled by President Joe Biden in October 2022, may become a law in the future. \url{https://www.whitehouse.gov/ostp/ai-bill-of-rights/}}, guidelines to protect user privacy have greatly strengthened, and will continue to do so. 

Of the many responses from the research community, synthetic data is seen as a solution to some of these privacy concerns. There are numerous synthetic data generation (SDG) techniques. Many of them aim to maintain important statistical properties of the real data that would make it useful, while generating the fake data with randomness to sever any connection to personally identifiable information.

However, because synthetic data is a relatively new technology (compared to artificial intelligence for example, which has sceptics enough in its own right), it has yet to gain significant adoption in industry \cite{jordon2022synthetic}. 

The explanation for this, in part, is that synthetic data may still leak private information about the training data, depending on the generation mechanism and the ``privacy budget'' used. For this reason, academically-motivated attacks have been developed to expose these vulnerabilities, and to empirically show which generation techniques protect privacy best.

One fundamental attack is the membership inference attack (MIA). The goal of membership inference is to determine which real data examples were used to train the SDG algorithm that produced the synthetic examples. For example, if synthetic data is generated from a database of cancer patients, information that links an actual individual to the fake data is theoretically removed. But an MIA makes observations on the fake data that gives confidence that an individual was present in the training set, allowing the attacker to infer that the individual has cancer.

MIAs can be conducted in other contexts, where instead of attacking a synthetic dataset, the goal is to attack a machine learning model from its output \cite{7958568}. But in this paper, we focus on the former, known as ``Inference-on-Synthetic" attacks \cite{houssiau2022tapas}. 

Different MIAs can rely on a different set of assumptions for the threat model; some MIAs use a threat model where an adversary knows little. Others assume a generous amount of knowledge in the hands of the adversary (the motivation being to build stronger privacy protections against a stronger opponent). 

A threat model with ``the auxiliary data assumption'' admits knowledge of the population data $D_{aux}$ to the adversary \cite{houssiau2022tapas}, in addition to already having the synthetic data $D_{synth}$. A threat model with ``black-box knowledge of the generator'' admits knowledge of the SDG algorithm and its hyperparameters used. We omit a discussion on the ``white-box'' threat model, where an adversary knows the internal weights and randomness used in the SDG algorithm. Always, the SDG algorithm's training data $D_{train}$ is hidden. 

Recent SDG algorithms, MST \cite{mckenna2021winning} and PrivBayes \cite{zhang2017privbayes}, have been shown to be strong against MIAs, while providing excellent utility. Importantly, unlike several other popular SDG algorithms, these both provide robust mathematical definitions of their privacy using \textit{differential privacy} (DP) \cite{dwork2014algorithmic}. DP is defined by 
$$
Pr[\mathcal{A}(D_1) \in O] \leq e^\varepsilon  Pr[\mathcal{A}(D_2) \in O],
$$
which bounds the probability that some arbitrary query $\mathcal{A}$ on two adjacent datasets $D_1, D_2$ of being equal by the parameter $\varepsilon$. Adjacent datasets are datasets that differ in only one record. ``$\varepsilon$-DP'' privacy is achieved when randomness within $\mathcal{A}$ is applied to the query in such a way that upholds this inequality. 

$\varepsilon$ can be selected in MST and PrivBayes to synthesize data that favors either privacy or utility (lower $\varepsilon$ means applying more noise, theoretically resulting in higher privacy). A high-level explanation of how MST and PrivBayes apply the randomness, and how they work in general, is given in the ``Our Approach'' section.

In this paper, we present two novel MIAs, one tailored specifically to MST and one to PrivBayes. These attacks are both based on the same underlying heuristic. We show that attacks using this heuristic are very successful for high values of $\varepsilon$, and thus demonstrate significant privacy leakage. Our approach makes both assumptions (i.e. knowledge of $D_{aux}$ and knowledge of the SDG algorithm used), and we discuss the merit in doing so. 

These attacks are both enhancements of the recently-proposed DOMIAS MIA \cite{van2023membership}. DOMIAS makes use of the auxiliary data assumption to great success, but doesn't assume knowledge of the SDG algorithm. Our work leverages this second assumption, and in so doing, strengthens the DOMIAS attack.

Our attacks are simple and efficient. While some MIAs require extensive computation of ``shadow'' synthetic datasets as a way of understanding the generator's behavior, the heuristic we use performs \textit{minimal} shadow modelling, and doesn't persist any shadow dataset. This is significant because burgeoning privacy laws consider ``[t]he practical feasibility of an attack... when assessing what constitutes anonymous data.'' \cite{hayes2017logan}. 

Lastly, our membership inference heuristic and experiments are motivated by our participation in the SNAKE 
(SaNitization Algorithm under attacK ...$\varepsilon$) 
Challenge\footnote{\url{https://dl.acm.org/doi/10.1145/3583780.3614754}} \cite{allard2023snake}, where our approach won first place. In this competition, we carry out an MIA on synthetic datasets generated using MST and PrivBayes, with $\varepsilon \in \{1, 10, 100, 1000\}$. The datasets are demographic in nature, containing fifteen ordinal and categorical features.

\section{DOMIAS Overview}

The novel MIA proposed recently, DOMIAS \cite{van2023membership}, outperforms many other approaches by leveraging the auxiliary data assumption. Conceptually, DOMIAS estimates the joint probability distribution of $D_{synth}$ (using some \textit{density} estimation $S$), and also of $D_{aux}$. With these, it simply divides the probability estimation of $D_{synth}$ by that of $D_{aux}$. 
$$\Lambda = \frac{S(D_{synth})}{S(D_{aux})}$$

The idea is that, if the SDG algorithm was at all overfit to its training data, then $S(D_{synth})$ should more closely resemble $S(D_{train})$ than $S(D_{aux})$. This discrepancy becomes pronounced in $\Lambda$. A member's presence may be inferred when $S(D_{synth})$ at that member's point is high relative to its probability at that point in $S(D_{aux})$.

\begin{figure}
\centering
\includegraphics[width=8.5cm]{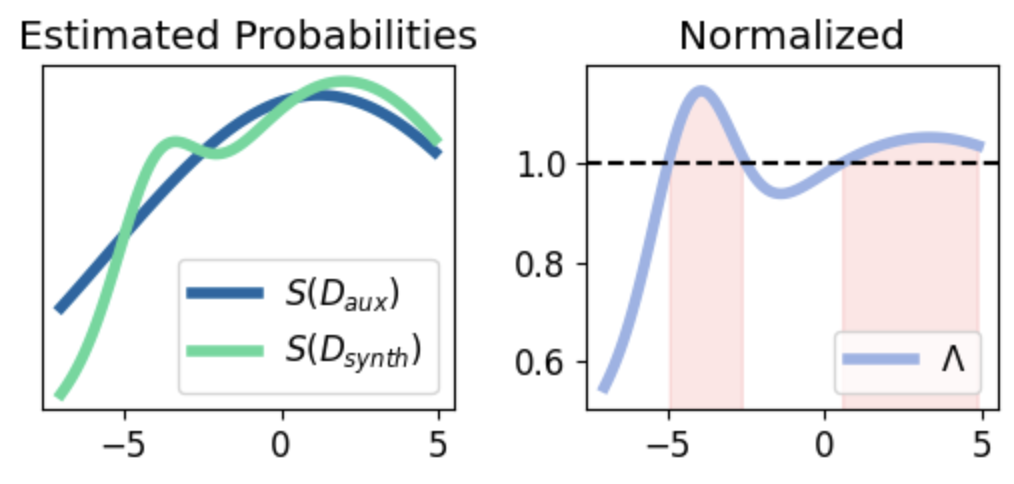}
\caption{A simple visualization of how DOMIAS detects overfitting. (left) $S$ gives an estimation of the probability distributions of $D_{synth}$ and $D_{aux}$. (right) Normalizing $S(D_{synth})$ by $S(D_{aux})$ exposes overfitting on $D_{train}$.} 
\label{fig:domias_example}
\end{figure}

Notice, in Figure \ref{fig:domias_example} (left), that if the estimated joint probabilities of $D_{synth}$ and $D_{aux}$ are depicted with curves, then normalizing the curve of $D_{synth}$ with $D_{aux}$ (right) makes clear where there may have been a concentration of training data used by the generator. In the figure, a candidate existing somewhere in the red shaded region are classified as a member by DOMIAS\footnote{In this example the threshold for membership is when $\Lambda = 1.0$. But it needn't be. Normalization allows for any threshold to identify overfitting meaningfully.}. Without the auxiliary data assumption, it's more difficult to contextualize observations on $D_{synth}$ and detect overfitting.

\section{Our Approach}

The success of DOMIAS largely depends on the quality of the density estimation. If poor, the estimated probability distributions won't distinguish $D_{synth}$ from $D_{aux}$. And in practice, when the datasets are large, key discrepancies between the two become overshadowed by noise. Since the DOMIAS attack assumes no knowledge of the SDG algorithm, it uses generic density estimations, such as KDE \cite{scott2015multivariate} and BNAF \cite{de2020block}. 

Our approach assumes black-box knowledge of the SDG algorithm. From this, we design two stellar density estimations to use with the DOMIAS attack, tailored to MST and PrivBayes. Our heuristic is to:
\begin{enumerate}
    \item \textit{identify} the way in which measurements taken on $D_{train}$ are chosen (we'll refer to these as ``focal-points'')
    \item \textit{simulate} the SDG algorithm several times (known as ``shadow modelling''), recording the frequency at which focal-points are chosen, thereby building a confidence of which focal-points were chosen when ran on ${D}_{train}$
    \item \textit{aggregate} these high-confidence focal-points as the density estimation, $S$
\end{enumerate}



To expound on the first step, MST and PrivBayes both approximate the joint probability distribution of $D_{train}$ in order to construct synthetic data that is meaningful. They do this by selecting several informative measures of $D_{train}$, which we call ``focal-points'', to retain in $D_{synth}$. They are chosen and measured differently in MST and PrivBayes, requiring our density estimation for each to be implemented slightly differently. 

These focal-points are highly relevant to an attacker because they show what the generators measured in $D_{train}$! The success of our density estimation depends on choosing the same ones the generator chose during its training because attaining such a measurement will give us the clearest possible picture of $D_{train}$. So determining how MST and PrivBayes choose them is the first step. 

Both algorithms are designed to choose focal-points that yield the highest information gain (or some variant thereof). But in order to adhere to $\varepsilon$-DP, both do so indeterminately.


This is where shadow modelling becomes useful. In the second step, we simulate the creation of $D_{synth}$ by running the same SDG algorithm several times, using the same $\varepsilon$ and training conditions. Since we do not know the true ${D}_{train}$, we using random samples of $D_{aux}$ as $\hat{D}_{train}$. From this process, we can record which focal-points are chosen, and with what frequency. This step involves obtaining an accurate implementation of the SDG algorithm, and modifying it to record the focal-points. No computation is necessary on any $\hat{D}_{synth}$, which are discarded. 

Since we are using all the same parameters, using random samples of $D_{aux}$, we can develop a confidence of which focal-points were chosen, and then measured by the SDG on the hidden ${D}_{train}$. For example, if we notice that MST chose to measure the marginal probability of feature-pair (`age', `income') 48 times out of 50 runs, we can say that the generator measured this marginal  from ${D}_{train}$ and maintained it in $D_{synth}$ with high likelihood.


These frequencies bring the true probability of which focal-points are selected by the generator into focus. Once the oft-chosen are known, we can build on any dataset a probability estimation that resembles the generator's estimation of the same data. 
But there are of course challenges to building this estimation. 

\subsubsection{Brief interpolation on inherent limitations}

For one, we can't know which focal-points were chosen by the generator with certainty because of the DP noise added during the decision process. So we weight our measurement of the focal-point by its frequency chosen in shadow modelling; if one focal-point is chosen only half the time, we neither include or exclude it in the density estimation, but rather cut its influence in half. This weighting is less important when $\varepsilon$ is large and the generator's focal-point-choosing becomes more deterministic.

Additionally, both MST and PrivBayes add DP noise on the measurements taken on the focal-points themselves. So the generator's probability estimation of $D_{train}$ will be quite inaccurate when $\varepsilon$ is small, and therefor creating a $D_{synth}$ that inherently reveals little about $D_{train}$.

And furthermore, even when $\varepsilon$ is arbitrarily large and the added noise is negligable (or if granted white-box access), the generator builds only an \textit{estimation} of the true joint probability distribution. Only constructing a perfect probability distribution will yield perfect information of $D_{train}$, which is computationally infeasable when $D_{train}$ is even moderately sized. 
\textbf{End interpolation}

\vspace{1em}

In the last step, we create the custom density estimation $S$ by aggregating measurements of the focal-points. How this aggregation is done is discretionary. This work achieved success by summing the weighted focal-point measurements from some $D$ at the values of a specific candidate. This is depicted in Algorithms \ref{alg:mst_density} and \ref{alg:privbayes_density}.

We take this result using $D_{synth}$, and then again using $D_{aux}$, for each of the candidates. And then we take the ratio between the two, $\Lambda$, as is done by DOMIAS.

This approach may be tailored to any synthetic data generator that estimates the probability distribution of a training dataset. We now describe how these focal-points are manifested in MST and in PrivBayes.

\subsubsection{MIA on MST}
MST \cite{mckenna2021winning} builds a graphical approximation of the joint probability distribution of $D_{train}$ where the nodes (variables) are the features in $D_{train}$, the edges are the two-way marginal probabilities of the two features, and the graph is an undirected tree. These marginals are made differentially private. During synthesis, MST creates data samples in proportion with the probabilities of all of the DP marginals in the graph.

The edges chosen by MST (i.e. marginals) are the focal-points in MST our density estimation. How the graph is constructed and the edges are chosen is most interesting to us. MST attempts to draw edges that create a maximum spanning tree (hence ``MST'') based on the information gained by each feature-pair. But the attempt is inexact because of the differential privacy applied, making the edge choices stochastic.


We build the MST density estimation (Algorithm \ref{alg:mst_density}) by aggregating the high-confidence marginals, which were determined through shadow-modelling. We take the marginal probability of a candidate's value, for each of these marginals measured on the dataset, and compute a weighted sum of these (line 5). The weights $w \in W$ are simply the frequencies the respective marginal was chosen during shadow modelling for that $\varepsilon$.

\begin{algorithm}[tb]
\caption{Membership Inference Attack on MST}
\label{alg:mst_density}
\textbf{Input}: $D_{synth}$ the synthetic dataset, $D_{aux}$ the auxiliary dataset, $F^2$ a list of feature-pairs of length $n-1$, $n$ is the number of features in $D_{synth}$, a corresponding list of weights $W$ of each of the feature-pairs in $F^2$, $T$ a list of candidate records. \\
\textbf{Output}: $\Lambda$ the densities of $D_{synth} / D_{aux}$ for each $t \in T$
\begin{algorithmic}[1] 
\STATE Let $M^2_s = \{m_{s,1}, m_{s,2} ... m_{s,n-1}\}$ be the 2-way marginals of feature-pairs in $F^2$ measured on $D_{synth}$
\STATE Let $M^2_b = \{m_{b,1}, m_{b,2} ... m_{b,n-1}\}$ be the 2-way marginals of feature-pairs in $F^2$ measured on $D_{aux}$
\STATE Let $\Lambda$ = [\,]
\FOR{$t \in T$}
\STATE Let $\Lambda[t] = \sum_{i=1}^{n-1} w_i \cdot m_{s,i}(t) / m_{b,i}(t)$
\ENDFOR
\STATE \textbf{return} $\Lambda$
\end{algorithmic}
\end{algorithm}

\subsubsection{MIA on PrivBayes}
Our tailored MIA on PrivBayes synthetic data is very similar to our approach for MST. PrivBayes \cite{zhang2017privbayes} also estimates the joint probability distribution of $D_{train}$ by constructing a graph. Except that, while MST constructs an undirected tree, with exactly $n - 1$ edges by default, PrivBayes constructs a \textit{directed} graph, where edges represent important \textit{conditional} probabilities between ``child'' features and sets of ``parent'' features.


PrivBayes \cite{zhang2017privbayes} also generates the synthetic data by referring to its graphical approximation of $D_{train}$\footnote{This is done by sampling values directly from the conditionals, and then following the corresponding path along the directed graph.}, making it highly relevant to an attacker. Like in MST, edges are preferred by PrivBayes that exhibit a variant of mutual information. But the choice of directed edges, and therefor the choice of which (and how many) parents each conditional has, are chosen stochastically, depending on $\varepsilon$. And lastly, DP-noise is added to the conditionals themselves. We use these conditionals as our focal-point for PrivBayes.


Since we don't know which conditionals were used to construct $D_{synth}$ we use shadow modelling to observe which conditionals are frequently selected for the graph creation. Using the correct $\varepsilon$ is especially important for PrivBayes, since not only does $\epsilon$ determine the variability of conditionals chosen, but it also determines which conditionals are \textit{preferred}.

Specifically, when $\varepsilon$ is small, PrivBayes reduces $k$, the maximum number of parents in its graphical approximation. Intuitively, setting $k$ to be large can allow conditionals to more closely approximate the true probability distribution. However, it also means that the high-specificity conditionals are more susceptible to the DP-noise. Throttling back $k$ mitigates this. So entirely different conditionals are tended towards for different $\varepsilon$s.


The computation of our density estimation is depicted in Algorithm \ref{alg:privbayes_density}. Here we depart from MST's ``feature-pair'' terminology, and use the more generic ``tuple'' to represent the features in a conditional. Since MST is building an acyclic tree (by default), its marginals only contain two features. On the other hand, PrivBayes conditionals can have any number of parent features, hence ``tuple''. We use the list of conditionals frequently chosen in shadow modelling to find the probability of a candidate instance's value for each of those conditionals, and then perform a weighted sum of those values (line 5) to derive a density at that candidate's location in the synthetic data and the auxiliary data. We use these densities for our modified DOMIAS measurement, $\Lambda$.

\begin{algorithm}[tb]
\caption{Membership Inference Attack on PrivBayes}
\label{alg:privbayes_density}
\textbf{Input}: $D_{synth}$ the synthetic dataset, $D_{aux}$ the auxiliary dataset, $F^2$ a list of feature-tuples of length $h$, a corresponding list of weights $W$ of each of the feature-tuples in $F^2$, $T$ a list of candidate records. \\
\textbf{Output}: $\Lambda$ the densities of $D_{synth} / D_{aux}$ for each $t \in T$
\begin{algorithmic}[1] 
\STATE Let $C_s = \{c_{s,1}, c_{s,2} ... c_{s,n-1}\}$ be the conditionals of feature-tuples in $F^2$ measured on $D_{synth}$
\STATE Let $C_b = \{c_{b,1}, c_{b,2} ... c_{b,n-1}\}$ be the conditionals of feature-tuples in $F^2$ measured on $D_{aux}$
\STATE Let $\Lambda$ = [\,]
\FOR{$t \in T$}
\STATE Let $\Lambda[t] = \sum_{i=1}^{h} w_i \cdot c_{s,i}(t) / c_{b,i}(t)$
\ENDFOR
\STATE \textbf{return} $\Lambda$
\end{algorithmic}
\end{algorithm}

\subsubsection{Activation Function}
Our goal is to get a probability, $P \in [0, 1]$, of the membership of each candidate. We convert $\Lambda$, which $\Lambda \in [0, \infty)$, into $P$ with the \textit{activation function}:
$$ P(\Lambda) = 1 / (1 + e^{-c (\log \Lambda - m)}) $$

In short, we take the logarithm of $\Lambda$ to convert the range to $(-\infty, \infty)$, and then pass it into the sigmoid function\footnote{Actually the range of the sigmoid function is $(0, 1)$, but with rounding happening towards the edges, it becomes $[0, 1]$.} to convert the range to $[0, 1]$. This maintains the monotonicity of $\Lambda$, where when $D_s / D_b = 1.0, P = 0.5$.

We also use sigmoid function parameters $c$ and $m$ in our approach, where $c$ is the confidence level, which determines how far from $0.5$ we want our probabilities, and $m$ is the median of all $\Lambda$s. By setting $m$ to the median, the function range is repositioned so that half of the $\Lambda$s are greater than $0.5$, and half are less. This is useful when an attacker expects half of the targets to be members. However, this parameter can be adjusted to divide the targets into any percentile. 

Or, in the more realistic setting, when an attacker has no knowledge or expectation of how many targets are members, setting
$P(\Lambda) =$ min$(\sqrt[c]{\Lambda}/2$, 1) can be a useful mapping of $\Lambda$ to probabilities, with $c$ similarly acting as a confidence level. This function maintains that when $\Lambda > 1$, probability of membership is $> 50\%$, which is consistent with the intuition behind how $\Lambda$ is defined, as discussed in ``DOMIAS Overview''.

DOMIAS and other works on MIAs stop short of defining an activation function, and score their results using AUC. We define ours here because of the good results it achieved in the SNAKE Challenge. But we also recognize that it may not be useful in real-world attack scenarios where the adversary doesn't know how many candidates are members to calibrate the function.



\section{Experimental Results}

\begin{figure}[t]
\centering
\begin{tabular}[b]{c  cc}
   $\varepsilon$ & MST & PrivBayes \\ \hline
   \midrule
   1 &  0.56  &  0.53  \\
   10 &  0.72  &  0.64  \\
   100 &  0.77  &  0.88  \\
   1000 &  0.77  &  0.96  \\
   \\ 
   \\ 
    \\
    \qquad
\end{tabular}
\centering
\includegraphics[width=4.2cm]{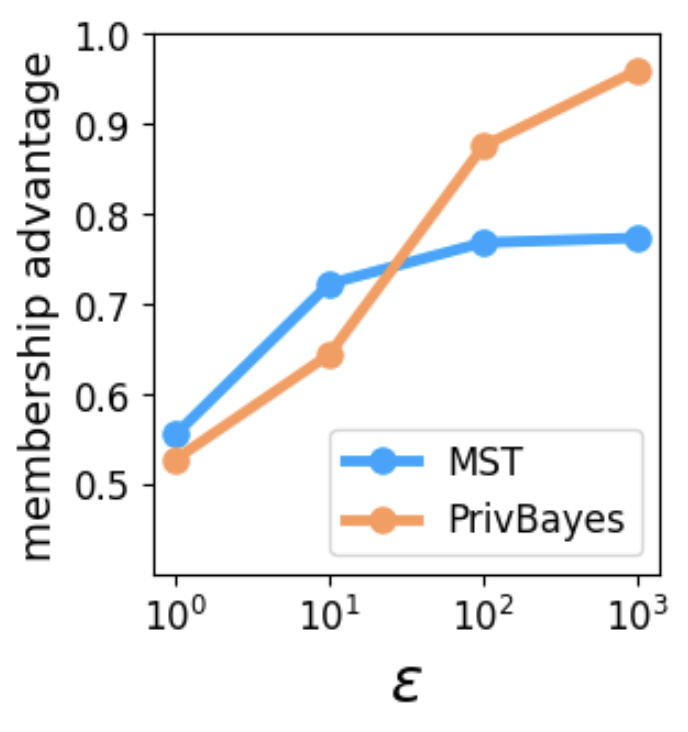}
\captionlistentry[table]{}
\caption{Membership advantage scores using our novel heuristic on different values of $\varepsilon$, averaged over 50 runs}
\label{fig:MA_scores}
\end{figure}

\subsubsection{Setup}

We conduct our experiments as defined by, and as part of our participation in the SNAKE Challenge \cite{allard2023snake}. Provided is an auxiliary dataset of 201,279 records, $D_{aux}$, each representing demographic information on an individual. There are 15 features, all represented discretely, five ordinal, and ten nominal. To give a sense, the features include age, state of residency, number of children, marital status, ethnicity, gender, field of profession, weekly hours worked, etc. 

The records (individuals) are also grouped into ``households". Households have between one and ten individual records. Of the 201,279 auxiliary records, there are 77,111 households. Instead of performing membership inference on individuals, the goal is to perform membership inference (MI) on entire households, called ``set MI" \cite{hilprecht2019monte}.

We conduct experiments on data generated by MST and PrivBayes, $D_{synth}$, and for value of $\varepsilon = \{1, 10, 100, 1000\}$. The records in the synthetic datasets are not grouped into households. These all contain 10,000 synthetic records.

For each of these eight experiments, each SDG is given a training set $D_{train}$. We sample one hundred ``candidate" households from the auxiliary data on which to perform set MI, $C$. A candidate household must contain at least five records. So, in practice, $C$ contains at least 500 records. We then sample 10,000 records from the auxiliary data that are not candidates; $D_{\overline{C}} \subseteq D_{aux} \setminus C$. A random sample of fifty households $M \subseteq C$ are appended to the auxiliary sample for training the synthetic dataset, $D_{\overline{C}} \cup M = D_{train}$.

We then measure the accuracy of our MIA on the synthetic data, using knowledge of $D_{aux}$ and which SDG algorithm and parameters were used. We perform each experiment 50 times. For each run, we generate a new training dataset $D_{train}$, a new candidate set $C$, and a new member set $M \subset C$, which are used in both the MST and PrivBayes experiments to maintain consistency. We score the \textit{membership advantage} (MA) on our predictions for each of the one hundred candidate households in $C$ against the ground truth $M$, and average the membership advantage across all fifty runs.

\subsubsection{Membership Advantage}
In alignment with the SNAKE challenge, we evaluate our predictions $P$ of the candidate household against the solutions using ``membership advantage'' \cite{yeom2018privacy}, defined as:
$$\textit{MA} = (\textit{tpr} - \textit{fpr} + 1) / 2$$
where $\textit{tpr}$, $\textit{fpr}$ are the true positive rate and false positive rate, weighted by $2 \cdot |0.5 - p_i|, p \in P$.

We predict the probability of membership for each \textit{household}. So this metric is used to evaluate, not predictions for individual records, but for membership of an entire household. Of the possible ways to make a household prediction, we make predictions for each individual record in the household (using our activation function with the threshold set to $0.5$), then simply take the average. This scored better than other approaches, such as trusting the most confident individual prediction of the  household, or trivial weightings therein.

\subsubsection{Results}

\begin{figure}
\vspace{0.5em}
\centering
\includegraphics[width=8.5cm]{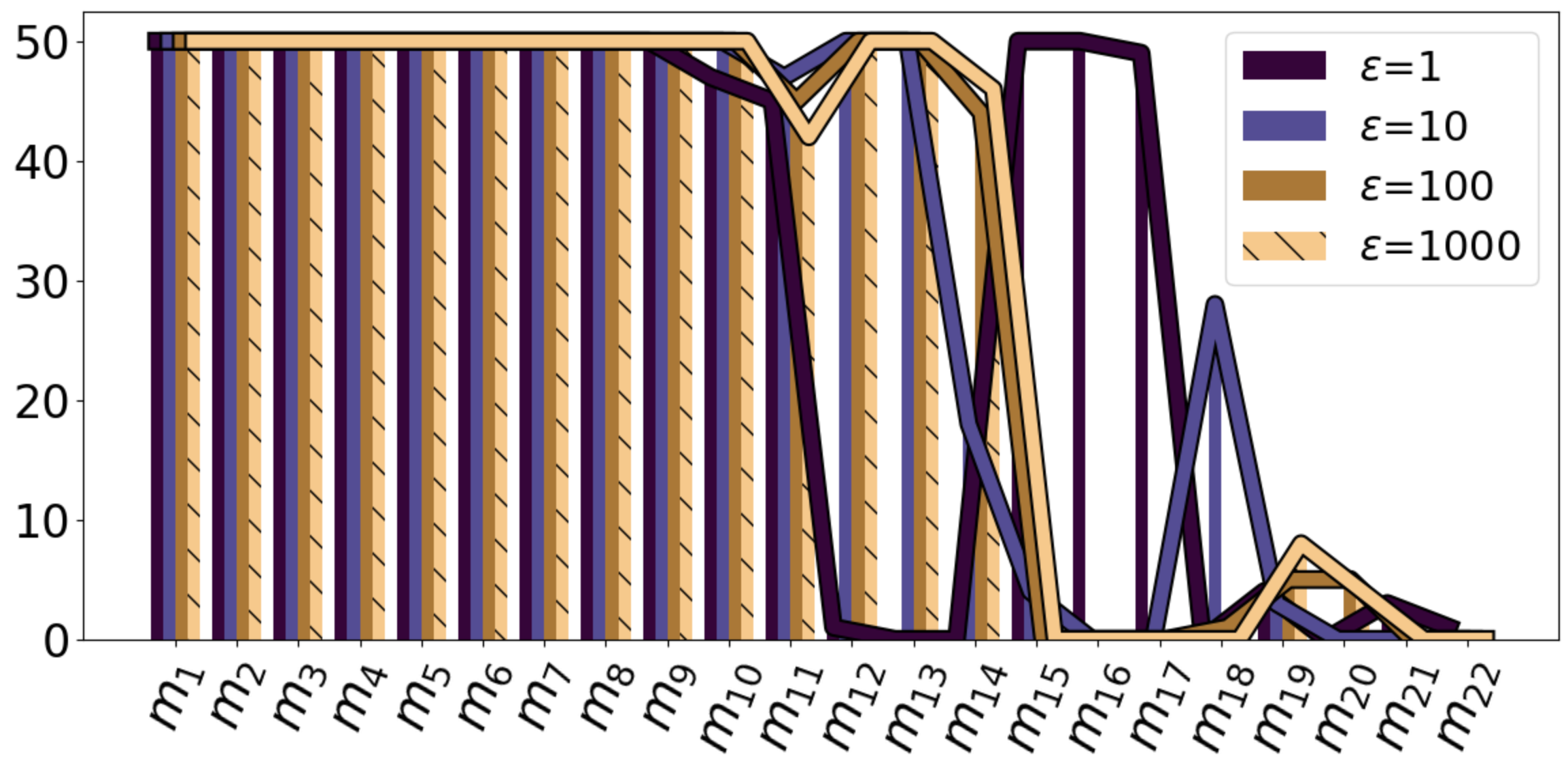}
\caption{Frequencies at which \textbf{marginals} are selected, when shadow modelling MST across 50 runs, for various $\varepsilon$} 
\label{fig:marginals_variability}
\end{figure}

\begin{figure}
\centering
\includegraphics[width=8.5cm]{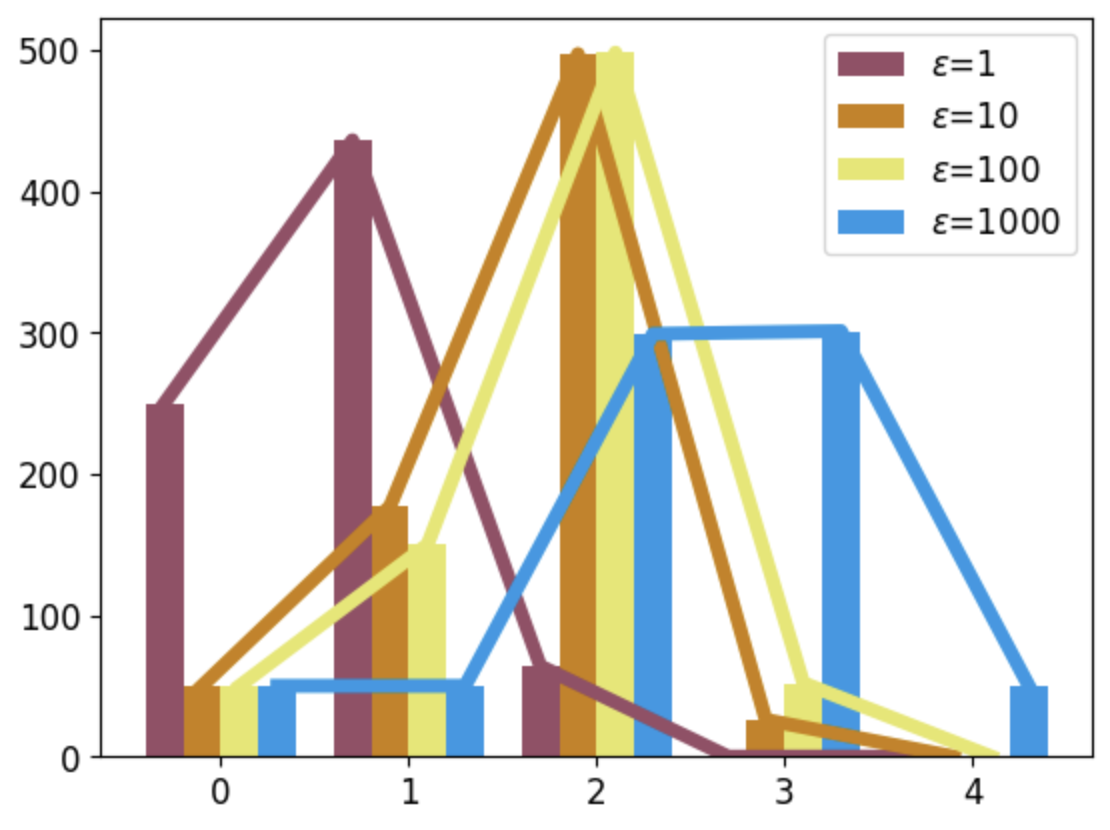}
\caption{Frequencies at which conditionals' \textbf{parent sizes} are selected in PrivBayes when shadow modelling across 50 runs, for various $\varepsilon$} 
\label{fig:conditionals_variability2}
\end{figure}

For both of our attacks on MST and PrivBayes, we achieved unambiguous increases in accuracy as $\varepsilon$ increases. As shown in Figure \ref{fig:MA_scores}, our attacks score an $\textit{MA}$ of 0.56 and 0.53 respectively, when $\varepsilon = 1$. When $\varepsilon = 1000$, our inference of members is very high, scoring 0.77 and 0.96.

Since these results were better than anticipated, we looked more closely into the source of our confident predictions. In our investigation, it became apparent that choosing which marginals or conditionals (the focal-points) were measured in our MST- and PrivBayes-tailored density estimation $S$ was critical to the attack's success.

We observed over 50 shadow runs, for $\varepsilon \in \{1, 10, 100, 1000\}$ that, as expected, the variability in which edges were chosen had an inverse correlation with $\varepsilon$; when $\varepsilon$ is high, the edges chosen in each shadow run were fairly consistent. However, when $\varepsilon$ decreased, there was greater variability in which edges were selected. We also note that, while variability in edge selection increased, the \textit{preference} of which edges were selected also shifted. This is expected for different $\varepsilon$ in PrivBayes, as we discuss below, but it remains puzzling behavior for MST.

For MST, we visualize this result as a bar graph in Figure \ref{fig:marginals_variability}. This shows that, of the 105 possible two-way marginals for $n=15$ features, $(n - 1)^2 / 2$, only 22 were selected at least once during all 200 runs (we name them arbitrarily $\{m_1, ... m_{22}\}$). Note that MST selects 14 edges (two-way marginals) each run to construct a tree on 15 features. $\{m_1, ... m_9\}$ were chosen every run, for every $\varepsilon$. Some of the variability in the remaining 13 edges chosen can be explained by the DP noise used during graph construction. But observe that $m_{15}, m_{16}, m_{17}$ were selected with high frequency \textit{only} when $\varepsilon = 1$. Yet those three edges weren't chosen at all when $\varepsilon = 100, 1000$. 

We make similar observations for PrivBayes, but present them from a different angle to recognize the difference between the graphs constructed by MST and PrivBayes. Figure \ref{fig:conditionals_variability2} shows that the conditionals chosen when $\varepsilon = 1$ have only ever 0, 1, or 2 parents. Alternatively, when $\varepsilon = 1000$, the graph constructed allows and favors the nodes having more parents. This translates to conditionals with greater specificity, more closely approximating the probability distribution of $D_{train}$. This behavior is expected (since we know that PrivBayes reduces the maximum number of parents $k$ with $\varepsilon$ to mitigate how extensively the DP noise is distributed). But this shows how differently sets of conditionals are chosen when $\varepsilon$ changes.

These findings show that the edge-choosing behavior of MST and PrivBayes is $\varepsilon$-dependent. So this underscores the importance of 1) using a set of focal-points in the attack's density estimation that resembles the ones used in construction as closely as possible, and 2) using shadow modelling with the correct value of $\varepsilon$ to predict what these focal-points are.

For the SNAKE Challenge, we were given the auxiliary data $D_{aux}$, and for each $\varepsilon \in \{1, 10, 100, 1000\}$, one target synthetic dataset $D_{synth}$ generated from MST, and one from PrivBayes, along with one set of candidates $D_{targets}$ from $D_{aux}$ on which to infer membership. Our predictions using this novel attack achieved the highest MA scores for all eight tasks (shown in Tables \ref{tab:snake_scores_mst} and \ref{tab:snake_scores_priv}), resulting in our team winning the final phase of the competition.

\begin{table}[H]
\centering
\begin{tabular}[b]{ c | cccc }
    $\varepsilon$    &  1  &  10  &  100  &  1000\\
    \hline
    \hline
    \textbf{Golob et al.}      & \textbf{0.61} & \textbf{0.79} & \textbf{0.76} & \textbf{0.81} \\
    (participant \#2) &  0.60  &  0.60  &  0.55  &  0.69  \\
    (participant \#3) &  0.60  &  0.53  &  0.56  &  0.54   \\
\end{tabular}
\caption{\textbf{MST} podium MA results for SNAKE Challenge }
\label{tab:snake_scores_mst}
\end{table}

\begin{table}[H]
\centering
\begin{tabular}[b]{ c | cccc }
    $\varepsilon$    &  1  &  10  &  100  &  1000\\
    \hline
    \hline
    \textbf{Golob et al.}      & \textbf{0.65} & \textbf{0.71} & \textbf{0.83} & \textbf{0.94} \\
    (participant \#2) &  0.53  &  0.62  &  0.69  &  0.51  \\
    (participant \#3) &  0.57  &  0.59  &  0.61  &  0.55   \\
\end{tabular}
\caption{\textbf{PrivBayes} podium MA results for SNAKE Challenge }
\label{tab:snake_scores_priv}
\end{table}

\section{Discussion}

This work shows that high-accuracy membership inference can be achieved against MST and PrivBayes synthetic datasets when $\varepsilon$ is large. The question of choosing $\varepsilon$ is dependent on a multitude of factors, including the desired utility and characteristics of the application's dataset. But care must be taken when the application's privacy concerns are significant. This work also shows how an adversary can use black-box knowledge of the SDG algorithm and which $\varepsilon$ used in an attack, if this information is published.

We conclude with a discussion on some of the questions raised by our approach. For one, why does the MA of our attack against MST taper off at around 0.8 for increasing $\varepsilon$, while it gets very close to 1.0 on PrivBayes? As Figure \ref{fig:conditionals_variability2} shows, PrivBayes with $\varepsilon = 1000$ chooses mostly several-parented conditionals in its graphical estimation of the joint probability distribution of $D_{train}$. This means that measurements of conditional probabilities of candidate values are highly specific. It also means that, if we happen to choose the same conditionals in our density function, then we will have measured exactly what PrivBayes measured on its train data, minus the unknown DP noise added. This, coupled with the DP noise being smaller, leads to a strong estimation of the density of $D_{train}$. 

On the other hand, MST only ever selects two-way marginals by default. So even if we choose the same exact set of marginals for our density estimation that MST chose when constructing the synthetic data, the specificity of these marginals can only approximate $D_{train}$ so well.

Our experiments shown in Figure \ref{fig:marginals_variability} beg the question of why MST favors marginals when $\varepsilon = 1$ that aren't selected when $\varepsilon$ is larger? Similarly, why does the \textit{consistency} of chosen marginals when $\varepsilon = 1000$ resemble the consistency when $\varepsilon = 1$? Across 50 runs, when $\varepsilon \in \{100, 1000\}$, MST chooses the same 14 marginals at least 42 times. But surprisingly, this is also the case when $\varepsilon = 1$. This oddity may be coincidental, and resolved with more exhaustive simulation.

Off the cuff, there is one defense MST can take; MST provides the option to manually select $n$-way marginals for its estimation of the training data's probability distribution. These hand-picked marginals obviously cannot be determined through shadow-modelling, which only simulate default behavior, and so would weaken an attacker's ability to build a similar estimation.

Since the field of synthetic data generation is so new, broader questions of this work can be raised. Whether or not the black-box knowledge assumption is reasonable is a well-founded one. However, we strongly believe that allowing generous knowledge to an adversary motivates the effort to build strong defenses -- an effort in which we hope to partake. 

Similarly, the high accuracies achieved in this paper could be questioned if $\varepsilon = 1000$ were considered excessive, or impractical. Given how large these datasets are, and how noisy, we don't consider using $\varepsilon = 1000$ out of the realm of possibility. Also, the point of our work is not to build a practical attack, but to bring potential vulnerabilities to light. A broader investigation into practical and theoretical SDG techniques will lead to more insight on this.

We are excited to continue pursuing this topic further. This includes conducting these experiments on DOMIAS proper to observe the empirical improvement in accuracy our attack yields. It includes experimenting on new datasets with varying domains, designing new density functions to exploit traits of other state-of-the-art synthetic generators like PATE-GAN, exploring $\varepsilon$'s effect on utility for different synthetic data generators, and how our attack's success on each generator calls for a reevaluation of that trade-off. 

And most importantly, future work includes improving or developing new synthetic data generation techniques that are resistant to membership inference attacks that use this heuristic.

\section{Acknowledgments}
Steven Golob and Sikha Pentyala are Carwein-Andrews Distinguished Fellows.
Steven Golob is supported by an NSF CSGrad4US fellowship. Sikha Pentyala is supported by the UW Global Innovation Fund and by a JP Morgan Chase PhD fellowship.

We would like to sincerely thank the organizers of the SNAKE challenge for posing this very interesting problem, and setting up an easy-to-use evaluation environment. Our research group, which has traditionally focused on privacy defenses, learned greatly from trying to break one. Looking at the chessboard from the other side can be invaluable.

\vspace{1em}
\bibliography{ref}

\end{document}